\documentclass[twocolumn,showpacs,preprintnumbers,amsmath,amssymb]{revtex4-1}
\usepackage{graphicx}
\usepackage{textcomp}

\begin{document}
\title{Many-body coherent destruction of tunneling in photonic lattices}
  \normalsize
\author{Stefano Longhi 
}
\address{Dipartimento di Fisica, Politecnico di Milano, Piazza L. da Vinci
32, I-20133 Milano, Italy}

%
\bigskip
\begin{abstract}
\noindent An optical realization of the phenomenon of many-body
coherent destruction of tunneling, recently predicted for
interacting many-boson systems by Gong, Molina and H\"{a}nggi [Phys.
Rev. Lett. {\bf 103}, 133002 (2009)], is proposed for light
transport in engineered waveguide arrays. The optical system enables
a direct visualization in Fock space of the many-body tunneling
control process.
\end{abstract}

\pacs{03.65.Xp, 42.82.Et, 32.80.Qk}


\maketitle

The phenomena of coherent destruction of tunneling (CDT)
\cite{Grossman91} and dynamic localization (DL) \cite{Dunlap86}
represent seminal results in studies of quantum dynamical control.
CDT and DL have been actively studied in several fields of physics
\cite{Grifoni}, culminating in recent experiments on strongly driven
Bose-Einstein condensates in optical lattices
\cite{Arimondo07,Oberthaler} which provided a very direct
observation of single-particle CDT and DL for matter waves.
Many-body generalizations of CDT have attracted considerable
interest in recent years, and experimentally
\cite{MB0,MB1,MB2,MB3,MB4,MB5}. CDT for many interacting particles
in a double well and its interplay with self trapping has been
discussed in \cite{MB0}, whereas shaking-induced renormalization of
the tunneling parameter and transition between superfluid and
insulator states was predicted and observed in Refs. \cite{MB1,MB5}.
In Ref.\cite{MB3}, Gong, Molina and H\"{a}nggi recently proposed a
novel route to CDT by considering a monochromatic fast modulation of
the self-interaction strength of a Bose-Einstein condensate in a
double well potential in the framework of a two-site Bose-Hubbard
model. Very interestingly, the modulation can be tuned such that
only an arbitrarily, a priori prescribed number of particles are
allowed to tunnel. In another physical context, the idea of CDT was
proposed \cite{Longhi05,nonlinearCDT} and experimentally observed
\cite{DellaValle07,Szameit09} for light waves in coupled-waveguide
structures, in which spatial propagation of light mimics the
temporal dynamics of a a driven quantum particle in a bistable
potential \cite{Longhi09LPR}. The optical analogues studied in
previous works \cite{Longhi05,nonlinearCDT,DellaValle07,Szameit09}
refer mostly to single-particle CDT. In the optical directional
coupler proposed in Ref.\cite{nonlinearCDT}, the addition of a Kerr
nonlinearity enables to mimic CDT in presence of particle
interaction in the mean-field limit, however such a scheme is not
capable of simulating the two-mode Bose-Hubbard model. In a recent
work \cite{Longhiun}, it was shown that a bosonic junction,
originally realized using Bose-Einstein condensates in a double well
potential \cite{Ob}, can be simulated using engineered waveguide
lattices. In this Report it is shown that modulation of the optical
lattice along the longitudinal direction enables to realize the
phenomenon of many-body CDT in a purely classical setting. As
compared to quantum simulators of the Bose-Hubbard dynamics, in
which the measurable quantities are usually the 'macroscopic' boson
number and phase difference in the two wells \cite{Ob}, the proposed
optical setting offers the rather unique possibility of a direct
visualization of the full many-body tunneling dynamics in Fock
space. Moreover, it offers the ability to prepare and observe small
systems of particle numbers, allowing to study the transition
between few-body and many-body behavior.
\par
 The starting
point of our analysis is provided by a standard two-mode
Bose-Hubbard Hamiltonian, which describes the tunneling dynamics of
bosons in a double-well potential (see, for instance, \cite{MB3,Ob})
\begin{equation}
\hat{H}(t)=\frac{\hbar v}{2}
(\hat{a}^{\dag}_l\hat{a}_r+\hat{a}^{\dag}_r \hat{a}_l)+\frac{\hbar
g(t)}{4} (\hat{a}^{\dag}_l\hat{a}_l-\hat{a}^{\dag}_r\hat{a}_r)^2,
\end{equation}
where $r$ (right) and $l$ (left) are the well sites, $\hat{a}_k$ and
$\hat{a}^{\dag}_k$ ($k=l,r$) are the bosonic annihilation and
creation operators, $v$ describes the constant tunneling rate
between the two modes, and $g(t)$ is the modulated interaction
strength between same-mode bosons. The total number of bosons
$N=\hat{a}^{\dag}_l\hat{a}_l+\hat{a}^{\dag}_r\hat{a}_r $ is a
conserved quantity and the dimension of the Hilbert space is
$(N+1)$. A sinusoidal modulation $g(t)=g_1 \sin(\omega t)$, without
dc bias, is assumed  for the interaction strength \cite{note0}. To
study the phenomenon of many-body CDT, in Ref.\cite{MB3} a Floquet
analysis of the Hamiltonian $\hat{H}(t)$ was performed after
introduction of the Schwinger representation of the angular momentum
operators. To provide an optical realization of the two-mode
Bose-Hubbard Hamiltonian and to understand the many-particle CDT
control scheme in an optical tunneling setup, we consider here the
Fock space representation of the Hamiltonian $\hat{H}(t)$. If we
expand the vector state of the system $|\psi(t) \rangle$ on the
basis of Fock states with constant particle number $N$, i.e. after
setting
\begin{equation}
| \psi(t) \rangle= \sum_{l=0}^N
\frac{c_l(t)}{\sqrt{l!(N-l)!}}\hat{a}^{\dag \; l}_{1} \hat{a}^{\dag
\; N-l}_{2} |0 \rangle
\end{equation}
the evolution equations for the amplitude probabilities $c_l(t)$ to
find $l$ bosons in the left well and the other $(N-l)$ bosons in the
right well, as obtained from the Schr\"{o}dinger equation $i \hbar
\partial_t | \psi(t) \rangle=\hat{H} | \psi(t) \rangle$, read
explicitly \cite{FP}
\begin{equation}
i \frac{dc_l}{dt}=(\kappa_lc_{l+1}+\kappa_{l-1}c_{l-1})+g(t) V_l c_l
\end{equation}
($l=0,1,2,...,N$), where we have set
\begin{equation}
\kappa_l=\frac{v}{2} \sqrt{(l+1)(N-l)} \; , \; V_l=\frac{1}{4}
(2l-N)^2.
\end{equation}
The normalization condition $\sum_{l=0}^N |c_l(t)|^2=1$ holds. The
evolution equations (3) for the Fock state amplitudes $c_l$ can be
viewed as formally analogous to the coupled-mode equations
describing light transport in a tight-binding array composed by
$(N+1)$ waveguides with longitudinally-modulated propagation
constant shift $g(t) V_l$ and engineered coupling rates $\kappa_l$
between adjacent waveguides, in which the temporal evolution of the
Fock-state amplitudes of the Bose-Hubbard Hamiltonian is mapped into
the spatial evolution of the modal field amplitudes of light waves
in the various waveguides along the axial direction $t$
\cite{Longhiun}. The fractional light power distribution $|c_l|^2$
in the various waveguides of the array at the propagation distance
$t$ thus reproduces the distribution of occupation probabilities of
the bosons between the two wells of the bistable potential. It is
worth noticing that, in the absence of particle interaction, i.e.
for $g=0$, the lattice model (3) associated to the Bose-Hubbard
Hamiltonian was previously introduced in the photonic context to
realize exact spatial beam self-imaging in finite waveguide arrays
\cite{Gordon} and shown to belong to a rather general class of
exactly-solvable self-imaging tight-binding lattices with
equally-spaced energy levels \cite{Longhi10}. In the high-frequency
modulation regime, the many-body CDT control method suggested in
Ref.\cite{MB3} can be simply explained as a tunneling inhibition
process between certain waveguides in the array. After introduction
of the slowly-varying amplitudes $a_l=c_l \exp[-i \int_0^t d \xi
g(\xi)]$, application of the averaging method \cite{Longhi08} yields
the following evolution equations for the amplitudes $a_l$
\begin{equation}
i \frac{da_l}{dt}=(\sigma _l a_{l+1}+\sigma_{l-1}^{*} a_{l-1})
\end{equation}
where we have set
\begin{equation}
\sigma_l=\kappa_l \langle \exp \left[ (V_l-V_{l-1})\int_0^t d \xi
g(\xi) \right] \rangle
\end{equation}
\begin{figure}
\includegraphics[scale=0.42]{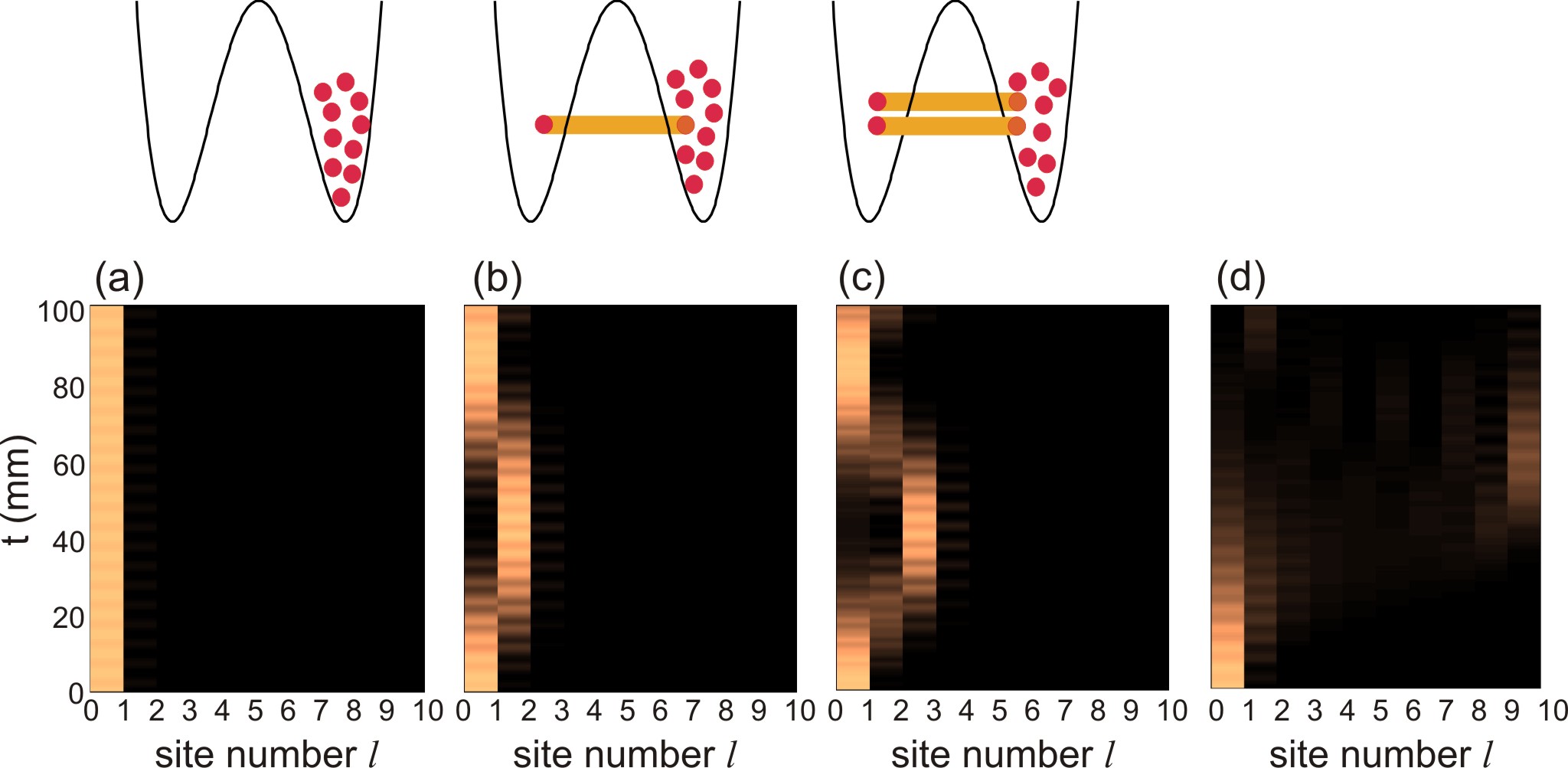}
\caption{(Color online) Evolution of the occupation probabilities
$|c_l(t)|^2$ of $N=10$ bosons for the two-site Bose-Hubbard
Hamiltonian (1) for $v=0.08 \; {\rm mm}^{-1}$, $\omega=0.628 \; {\rm
mm}^{-1}$ and for a few values of the modulation amplitude $g_1$:
(a) $g_1=1.679 \times 10^{-4} \; {\rm mm}^{-1}$; (b) $g_1=2.159
\times 10^{-4} \; {\rm mm}^{-1}$, (c) $g_1=3.022 \times 10^{-4} \;
{\rm mm}^{-1}$, and (d) $g_1=1.346 \times 10^{-4} \; {\rm mm}^{-1}$.
The four values of the modulation amplitude $g_1$ correspond to the
four vertical lines I-IV in the quasi-energy diagram of Fig.2(b).
The modulation amplitudes in (a), (b) and (c) correspond to
selective tunneling of zero, one and two bosons, respectively, as
schematically indicated in the panels at the top of the figures.}
\end{figure}
and the bracket stands for the time average over one oscillation
cycle. Equations (5) and (6) provide a first-order approximation to
the exact tunneling dynamics in the lattice (for more details see
\cite{Longhi08}). For a sinusoidal modulation $g(t)=g_1 \sin (\omega
t)$ and using Eq.(4), the explicit form of the effective coupling
rate $\sigma_l$ reads
\begin{equation}
\sigma_l=\kappa_l \exp(i \varphi_l) J_0 \left(
\frac{g_1(N-2l-1)}{\omega} \right)
\end{equation}
where $J_0$ is the Bessel function of order zero and
$\varphi_l=(g_1/ \omega)(2l-N-1)-l \pi/2$ is an inessential phase
term.
\begin{figure}
\includegraphics[scale=0.48]{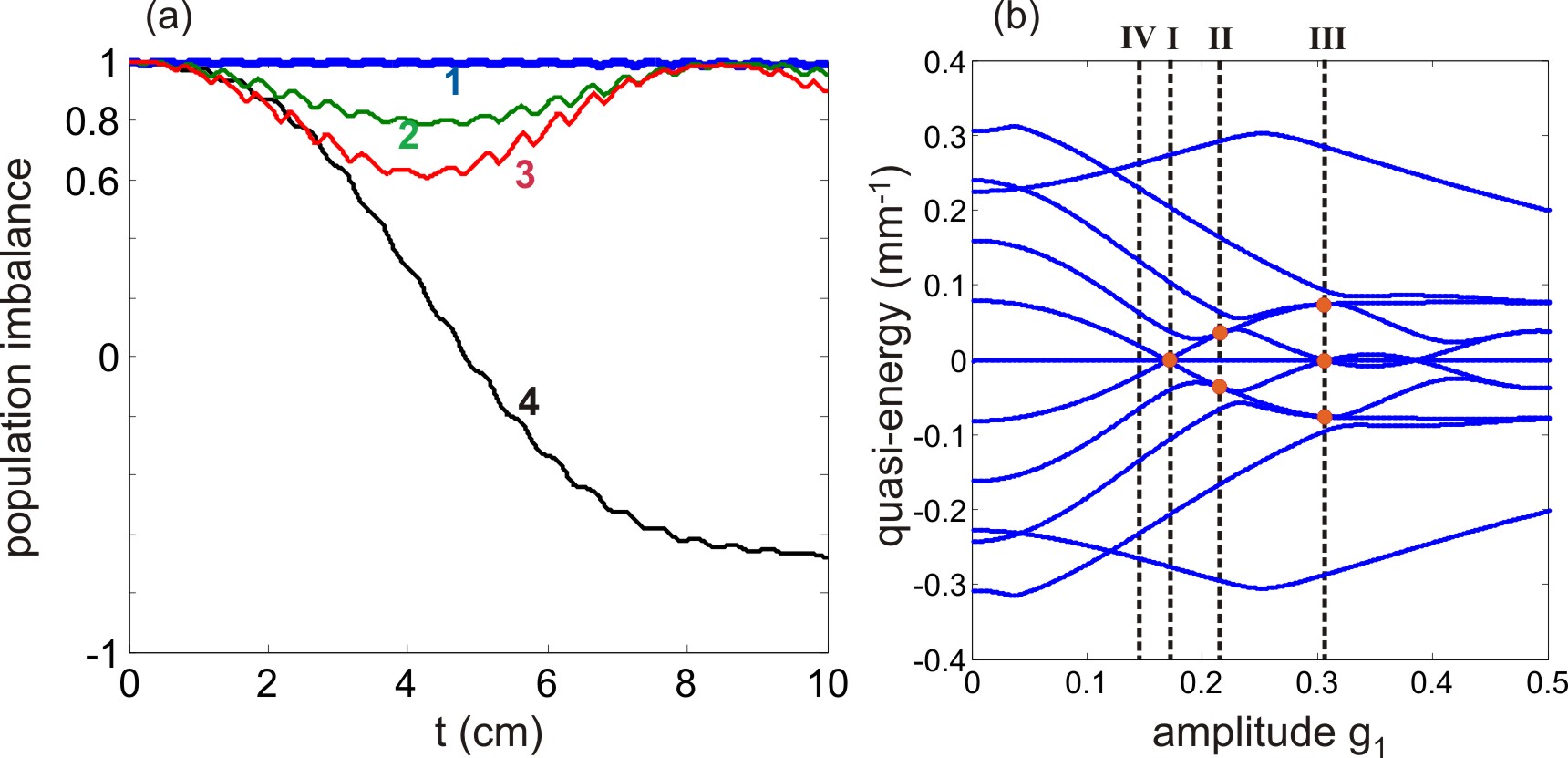}
\caption{(Color online) (a) Evolution of the normalized population
imbalance $S(t)$ corresponding to the simulations shown in Fig.1
(curve 1: $g_1=1.679 \times 10^{-4} \; {\rm mm}^{-1}$; curve 2:
$g_1=2.159 \times 10^{-4} \; {\rm mm}^{-1}$; curve 3: $g_1=3.022
\times 10^{-4} \; {\rm mm}^{-1}$; curve 4: $g_1=1.346 \times 10^{-4}
\; {\rm mm}^{-1}$). (b) Quasi-energy spectrum of the two-site
Bose-Hubbard Hamiltonian (1) versus $g_1$ for $v=0.08 \; {\rm
mm}^{-1}$ and $\omega=0.628 \; {\rm mm}^{-1}$. The vertical lines I,
II and III correspond to selective tunneling of zero, one and two
bosons, respectively, as shown in Figs.1(a), (b) and (c).}
\end{figure}
Suppose now that the system is initially prepared with all the $N$
bosons on the right well, i.e. $c_l=\delta_{l,0}$. According to
Ref.\cite{MB3}, a desired number $l_0=0,1,2,...$ of particles can be
allowed to tunnel to the left well provided that the amplitude $g_1$
of the modulation is chosen in such a way that
$g_1(N-2l_0-1)/\omega$ is a root of the $J_0$ Bessel function. In
this case, since the effective hopping rate $\sigma_{l_0}$ between
the sites $l_0$ and $l_0+1$ vanishes, one has $c_l(t)=0$ for $l \geq
l_0+1$, and tunneling to the lattice sites $l=l_0+1,l_0+2,...,N$ is
thus forbidden. Since $|c_l(t)|^2$ is the probability to find $l$
bosons in the left well and the other $(N-l)$ bosons in the right
well, this means that no more than $l_0$ particles can tunnel from
the right to the left well. As discussed in Ref.\cite{MB3}, the
conditions for selective CDT correspond to the crossing or touching
of quasi-energy Floquet states of Eq.(3) with opposite parity. In
our optical realization of the driven Bose-Hubbard Hamiltonian, the
onset of many-body CDT is thus simply visualized as the inhibition
of light tunneling in a portion of the array. As an example,
Figs.1(a-c) show the evolution of the occupation probabilities
$|c_l(t)|^2$, numerically computed from Eq.(3), corresponding to
tunneling of zero, one and two particles for a $N=10$ bosonic
system, initially prepared with all the particles in the right well,
for parameter values $v=0.08 \; {\rm mm}^{-1}$, $\omega=0.628 \;
{\rm mm}^{-1}$ and for the three different values $g_1$ of the
modulation satisfying the selective CDT conditions
$g_1(N-2l_0-1)/\omega=2.405$ with $l_0=0,1$ and 2. The corresponding
evolution of the normalized population imbalance \cite{MB3}, given
by $S(t)=\sum_{l=0}^{N} [(N-2l)/N] |c_l(t)|^2$, is depicted in
Fig.2(a). The selective CDT realized in the three cases corresponds
to the three points I, II and III in the quasi-energy diagram shown
in Fig.2(b). For comparison, in Fig.1(d) and curve 4 of Fig.2(a) the
behavior of the occupation probabilities and normalized population
imbalance is also shown for a wrong value of the modulation
amplitude $g_1$, which does not correspond to degeneracy of Floquet
states [see line IV in Fig.2(b)]. Note that, as opposed to
Figs.1(a)-(c), in this case the light waves spread over the entire
waveguides of the array.\par
 To realize the lattice model (3), the
hopping rates and site potentials should be tailored according to
Eqs.(4). In a waveguide array, this requires a control of waveguide
spacing and of refractive index contrasts and/or channel sizes of
the various waveguides. The tight-binding model (3) can be derived
using a variational procedure starting from the paraxial and scalar
wave equation for the electric field amplitude $\phi(x,t)$
describing the propagation of monochromatic light waves at
wavelength $\lambda$ in an array of $(N+1)$ waveguides with a
longitudinally-modulated refractive index profile $n(x,t)$ and
substrate index $n_s$
\begin{equation}
i \lambdabar \partial_t \phi= -\frac{\lambdabar^2}{2n_s}\partial_x^2
\phi+[n_s-n(x,t)] \phi,
\end{equation}
where $\lambdabar=\lambda/(2 \pi)$ is the reduced wavelength of
photons and $t$ is the longitudinal spatial coordinate. The hopping
rates $k_l$ and propagation constant shifts $gV_l$ entering in the
reduced equations (3) are given by certain overlapping integrals
involving the modal field and index profiles of the guides (for
details see, for instance, \cite{Longhi06PLA}). In particular, to
independently engineer the coupling constants $\kappa_l$ and
propagation constant shifts $g V_l$ of the waveguides, one can
assume a chain of waveguides with equal normalized refractive index
profile $G(x)$, but with different and longitudinally-modulated
index contrasts $\Delta n_l(t)$ ($l=0,1,2,...,N$) and spacing
$d_l=x_{l}-x_{l-1}$ ($l=1,2,...,N$), i.e. $n(x,t)-n_s=\sum_{l=0}^{N}
\Delta n_l(t) G(x-x_l)$.
\begin{figure}
\includegraphics[scale=0.68]{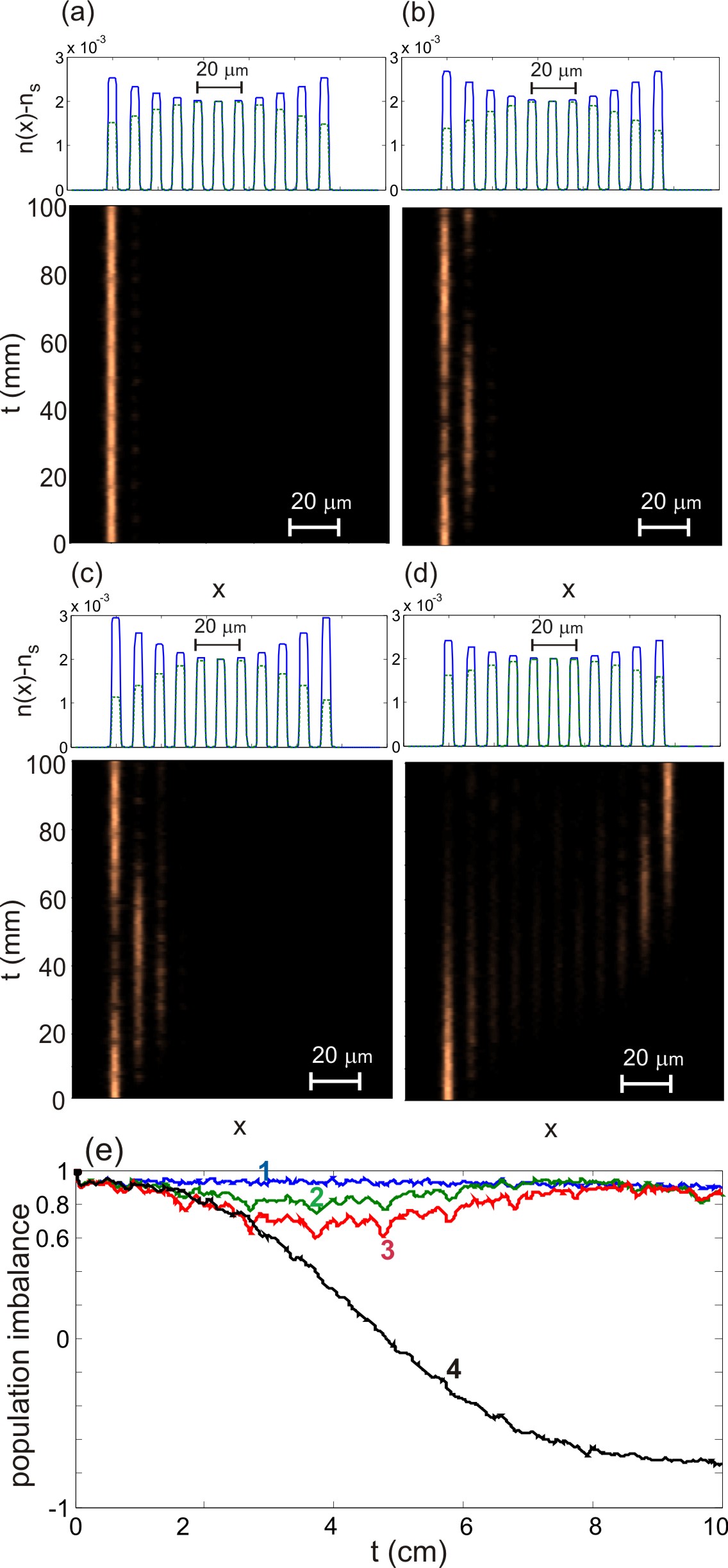}
\caption{(Color online) (a-d) Evolution of beam intensity (snapshot
of $|\phi(x,t)|^2$) in four 10-cm-long longitudinally-modulated
waveguide arrays, composed by $(N+1)=11$ waveguides, that realize
the lattice models of Figs.1(a-d). The refractive index profiles
$n(x)-n_s$ of the arrays, corresponding to the maximum/minimum of
modulation amplitude [$\sin(\omega z)= \pm 1$] are shown in the
upper panels of the figures by the solid/dashed curves. The behavior
of the normalized population imbalance $S(t)$, retrieved from the
beam center of mass position, is shown in (e) for the four arrays.}
\end{figure}
To realize the two-mode Bose-Hubbard lattice (3), the values of
$d_l$ and $\Delta n_l$ slightly vary around some mean values $d_r$
and $\Delta n$ that define a uniform array \cite{Longhiun}. The
waveguide separation $d_l$ mainly determines the value of the
coupling rate $\kappa_{l-1}$, with a characteristic exponential
dependence of $\kappa_{l-1}$ from $d_l$, whereas the index change
$\Delta n_l$ mainly defines the propagation constant mismatch $g_1
V_l$. Numerical simulations were performed to reproduce the
parameter conditions of Fig.1 for an operational wavelength
$\lambda=633$ and assuming $n_s=1.45$, which applies to fused silica
waveguide arrays. The normalized waveguide profile used in the
simulations is given by $G(x)=\{ {\rm erf}[(x+w)/D_x]-{\rm
erf}[(x-w)/D_x] \}/[2 {\rm erf} (w/D_x)]$, with channel width $2w=4
\; \mu$m and diffusion length $D_x=0.3 \; \mu {\rm m}^{-1}$. To
determine the distribution of distances $d_l$, a reference value
$\Delta n=2 \times 10^{-3}$ of refractive index contrast was
assumed, and correspondingly the coupling rate $\kappa$ between two
adjacent waveguides versus distance $d$ was computed, yielding to a
good approximation the exponential dependence $\kappa(d)=\kappa_0
\exp[-\gamma(d-d_r)]$ for distances not too far from the reference
distance $d_r=9 \; \mu$m, where $\kappa_0 \simeq 0.2144 \; {\rm
mm}^{-1}$ and $\gamma \simeq 0.6 \; \mu {\rm m}^{-1}$. The resulting
distance distribution that realizes the coupling rates (4) with
$v=0.08 \; {\rm mm}^{-1}$ indicates that $d_l$ varies in the range
9-9.916 $ \mu$m. By modulating the index contrasts $\Delta n_l$ of
the waveguides around the reference value $\Delta n$, four different
arrays were then designed to realize the four different interaction
regimes of Fig.1. A numerical computation of the propagation
constant of the channel waveguide indicates that a propagation
constant shift $g_1 \sin(\omega t) V_l$ is approximately obtained by
assuming a refractive index contrast $\Delta n_l=\Delta n+ \beta g_1
\sin(\omega t)  V_l \lambdabar$, where $\Delta n= 2 \times 10^{-3}$
is the reference value of the index contrast and $\beta$ is a
numerical factor of order one ($\beta \simeq 1.23$ for the chosen
error function waveguide profile). The resulting refractive index
profile of the arrays, corresponding to the maximum [$\sin(\omega
t)=1$] and minimum [$\sin(\omega t)=-1$] of the modulation cycle,
are depicted in the upper panels of Figs.3(a-d). Such arrays could
be fabricated in fused silica by the femtosecond laser writing
technique \cite{Szameit10}, in which the different refractive index
contrasts are obtained by varying the speed of the writing laser
beam. Figures 3(a-d) show the evolution of light intensity
$|\phi(x,t)|^2$ along the four arrays, as obtained by numerical
simulations of Eq.(8) using a standard pseudospectral split-step
method, when the left boundary waveguide is excited at the input
plane. The behavior of the normalized population imbalance $S(t)$,
which provides a clear signature of selective CDT \cite{MB3}, can be
retrieved from a measurement of the beam center of mass position
$\langle
 x(t) \rangle=\int dx x |\phi(x,t)|^2 / \int dx  |\phi(x,t)|^2$ using the simple relation $S(t)
\simeq 1-2\langle x(t) \rangle /(N d_r)$ \cite{Longhiun}. The
corresponding behavior of $S(t)$ for the four arrays is depicted in
Fig.3(e), which clearly indicates the onset of selective CDT
according to the results shown
in Figs.1 and 2.\\
In conclusion, a classic wave optics realization of the selective
many-body coherent destruction of tunneling, recently predicted for
interacting many-boson systems in Ref.\cite{MB3}, has been proposed
for light transport in modulated waveguide lattices. In such
photonic structures light propagation in the various waveguides maps
the evolution of the bosonic distribution in the double-well
potential, and thus enables to visualize the entire many-body
tunneling dynamics in the Fock space, a possibility which would be
hard to
achieve in matter wave systems.\\
\\
This work was supported by the Italian MIUR (Grant No.
PRIN-20082YCAAK, "Analogie ottico-quantistiche in strutture
fotoniche a guida d'onda").


\begin{thebibliography}{}
\bibitem{note0}
As discussed in Ref.\cite{MB3}, a dc bias of $g(t)$  can induce
population localization via a self-trapping mechanism. To isolate
population localization due to possible CDT phenomena from that due
to self-trapping, we will consider here solely an ac modulation.
\end{thebibliography}
\end{document}